\documentclass[10pt,twocolumn,letterpaper]{article}

\usepackage{iccv}
\usepackage{times}
\usepackage{epsfig}
\usepackage{graphicx}
\usepackage{amsmath}
\usepackage{amssymb}

\usepackage{caption}
\usepackage{color}
\usepackage{graphics,subcaption}
\usepackage{booktabs}
\usepackage{adjustbox}
\usepackage{multirow}
\usepackage{colortbl}
\usepackage{diagbox}
\definecolor{Gray}{gray}{0.9}
\usepackage{color, colortbl}
\usepackage[dvipsnames]{xcolor}
\usepackage{wrapfig}
\usepackage{floatrow}
\usepackage[T1]{fontenc}
\usepackage[utf8]{inputenc}
\usepackage[font=small,labelfont=bf]{caption}
\usepackage{graphicx}
\usepackage{pifont}%
\newcommand{\xmark}{\ding{55}}%
\usepackage{listings}
\usepackage{floatrow}
\usepackage{authblk}

\usepackage[pagebackref=true,breaklinks=true,letterpaper=true,colorlinks,bookmarks=False]{hyperref}
\hypersetup{
    pdfcreator={None},
    pdfproducer={None}
}

\DeclareMathOperator{\atantwo}{atan2}
\DeclareFloatSeparators{myfill}{\hskip.008\textwidth plus1fill}
\iccvfinalcopy 

\def\iccvPaperID{4000} 

\ificcvfinal\fi

\begin{document}

\title{The Devil is in the Upsampling:  Architectural Decisions Made Simpler for Denoising with Deep Image Prior}

\makeatletter
\def\thanks#1{\protected@xdef\@thanks{\@thanks
        \protect\footnotetext{#1}}}
\makeatother

\author[1*]{ \vspace{-0.3in} Yilin Liu}
\author[1*]{Jiang Li}
\author[1*]{Yunkui Pang \thanks{* These authors contribute equally.}}
\author[1]{Dong Nie}
\author[1]{Pew-Thian Yap \vspace{-0.16in}}
\affil[1]{University of North Carolina at Chapel Hill \tt\small \authorcr \{yilinliu, yunkuipa, dongnie\}@cs.unc.edu, jiang.li@unc.edu, ptyap@med.unc.edu}

\maketitle
\ificcvfinal\thispagestyle{empty}\fi

\begin{abstract}
   Deep Image Prior (DIP) shows that some network architectures inherently tend towards generating smooth images while resisting noise, a phenomenon known as spectral bias. Image denoising is a natural application of this property. Although denoising with DIP mitigates the need for large training sets, two often intertwined practical challenges need to be overcome: architectural design and noise fitting. Existing methods either handcraft or search for suitable architectures from a vast design space, due to the limited understanding of how architectural choices affect the denoising outcome. In this study, we demonstrate from a frequency perspective that unlearnt upsampling is the main driving force behind the denoising phenomenon with DIP. This finding leads to straightforward strategies for identifying a suitable architecture for every image without laborious search. Extensive experiments show that the estimated architectures achieve superior denoising results than existing methods with up to 95$\%$ fewer parameters. Thanks to this under-parameterization, the resulting architectures are less prone to noise-fitting\footnote{ \url{https://github.com/YilinLiu97/FasterDIP-devil-in-upsampling.git}}.
\end{abstract}

\section{Introduction}
Image denoising is useful on its own and can be a plug-in module for many other image restoration tasks \cite{zhang2017beyond,zhang2017learning,chan2016plug}. Deep neural networks have become the tool of choice for image denoising owing to their ability to learn natural image priors from large-scale datasets. 
Yet, Deep Image Prior (DIP) \cite{ulyanov2018deep} requires only a single degraded image for image resotration. Remarkably, DIP shows that a randomly initialized convolutional neural network (CNN) can regularize image restoration through its architecture and early-stopping optimization. 
This is inspired by the phenomenon that \textit{some} network architectures act inherently as image priors, favoring generating smooth, natural images and resisting noises or degradations. However, image denoising with DIP is greatly influenced by architectural design \cite{chakrabarty2019spectral,arican2022isnas,ulyanov2018deep,heckel2018deep,ho2021neural}, and the associated tendency to fit the original noisy image, i.e., over-fitting \cite{heckel2018deep,heckel2019denoising}.

\begin{figure}
\includegraphics[width=\columnwidth]{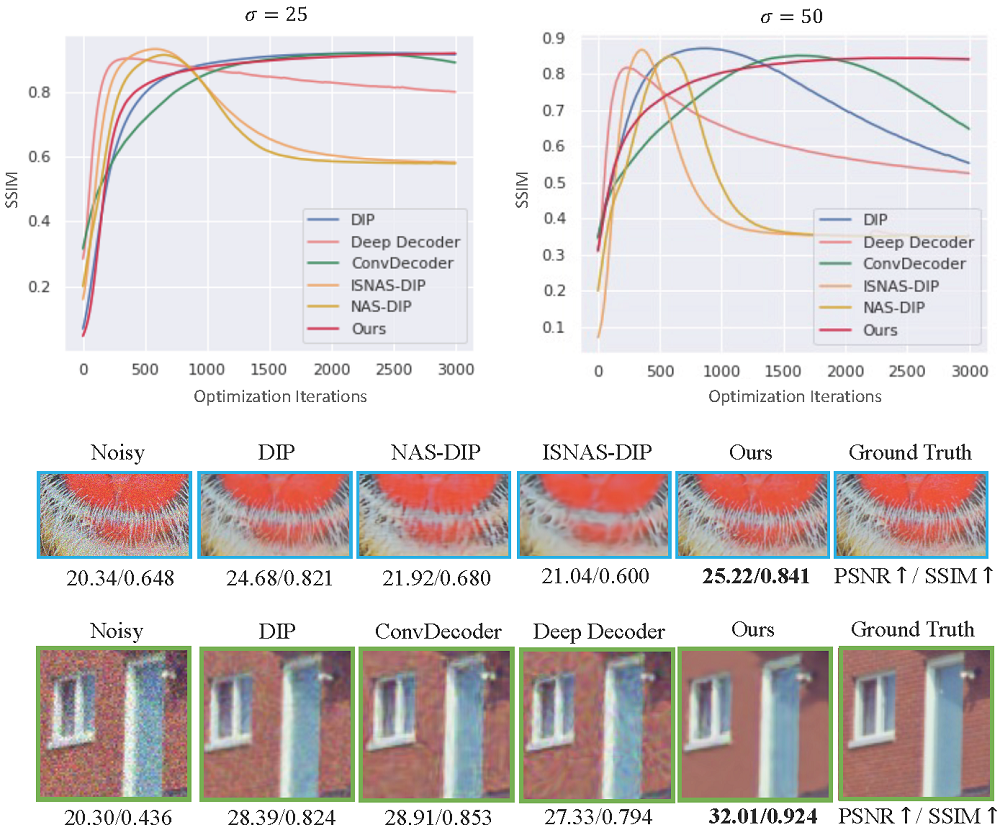} \label{fig:1}
\caption{\textbf{Top}: Performance (SSIM$\uparrow$) under different levels of Gaussian noises. \textbf{Bottom}: Denoising of a fine-grained ($1^{st}$row) and a coarse-grained image ($2^{nd}$row). Most existing methods, including the recent image-specific ISNAS-DIP\cite{arican2022isnas}, struggle to perform well simultaneously in both cases. Our simple strategies are flexible in image-specific architectural adaptation without requiring a search. Moreover, the results of the lightweight ConvDecoder\cite{darestani2021accelerated} and Deep Decoder\cite{heckel2018deep} suggest that without proper model setups, under-parameterization itself can neither ensure good denoising performance nor remove the need for early-stopping.}
\end{figure}

Architectural design for DIP remains an open problem. One prevailing view is that model under-parameterization limits noise over-fitting and thus mitigates the need for early stopping \cite{heckel2018deep}. However, our experiments reveal that multiple model architectures can exist under a similar parameter budget, where inappropriate model setups can still lead to noise fitting and over-smoothing (Fig.\hyperref[fig:1]{1}, Fig.\hyperref[fig:upsampling] {3}). Another line of work automates the architecture identification using Neural Architecture Search (NAS)\cite{chen2020dip,ho2021neural,arican2022isnas}. Without prior knowledge on suitable architectures, extensive search incurs substantial computational costs, prohibiting image-wise NAS for optimal restoration \cite{arican2022isnas}. Arican et al. \cite{arican2022isnas} narrow the search space using training-free metrics, but the need for candidate comparison dramatically prolongs the restoration time ($\sim7$ hours/image). Moreover, NAS-based models, along with many DIP models, are typically heavily parameterized and prone to over-fitting, as corroborated in our experiments. Thus, their performance largely depends on the timing of early stopping, which is typically image-specific and hard to pinpoint without access to ground truth.

Directly identifying an effective under-parameterized architecture for each image is challenging in light of the vast number of different architectures and the absence of ground truth for explicit supervision. To simplify the architectural decisions for DIP, we rethink the architectural influences on its performance in the context of image denoising.

We start by noting that denoising performance is attributable to \textit{only a few componenets}, primarily the unlearnt upsampling operations. Our frequency analysis reveals that the fixed upsampling operations tend to bias the architecture towards low-frequency contents more strongly than linear or convolutional layers, critically influencing both the peak PSNR and the point of early stopping to avoid noise-fitting.

Importantly, this finding leads to empirical discovery on the roles of typical architectural components in DIP: assuming a standard hourglass network, \textbf{i)} simply scaling the depth and width can balance smoothing and preservation of details, due to the low-pass filtering effects of the upsampling operations inserted in-between the layers. As we observed, a wider and shallower network is better at preserving details and therefore suitable for fine-grained images. This suggests that the "optimal" architecture can vary across images, and image texture should be considered for more effective denoising. \textbf{ii)} Skip connections make a deep network perform similarly as a shallower one likely by reducing the "effective upsampling rate". This implies the possibility of discarding skip connections to simplify DIP architectural design. 

Based on these insights, we find it sufficient to restrict the design choices to only the network depth and width, reducing the problem to searching through only a handful of sub-networks and adapting them according to the level of image details. This can be done as a pre-processing step without costly searching or evaluation. We show that this simple strategy works with both the hourglass and decoder structures, and with proper setups, the estimated networks can denoise while preserving the details better than the larger networks with only 5$\%$$\sim$40$\%$ number of parameters. The resulting under-parameterization alleviates the need for early stopping. Our contributions are as follows:
\begin{itemize}
    \item We pinpoint that unlearnt upsampling is the main driving force behind the spectral bias of DIP.
    \item Leveraging this insight, we empirically identify the influences of depth, width and skip connections, along with their correlations with image texture, allowing for \textbf{quick}, \textbf{effective} and \textbf{more interpretable} architectural design for \textit{every image} without the laborious search.
    \item We are the first to associate DIP architectural design with image texture. To promote future research, we build a \textit{Texture-DIP Dataset} consisting of images from three popular datasets, reclassified into several predefined width choices -- validated through experiments, according to textural complexity.
    \item We show that with proper setups, a highly under-parameterized subnetwork could match and even outperform larger counterparts, especially at a higher noise level. We conducted extensive experiments on synthetic and real-world noise to validate our findings and approach.
\end{itemize}

\section{Related Work}
\textbf{DIP Variants.} Deep Decoder \cite{heckel2018deep} is an under-parameterized network proposed to avoid early stopping. However, our investigations show that under-parameterization alone is \textit{not} sufficient for denoising. For instance, ConvDecoder \cite{darestani2021accelerated}, a convolutional variant of Deep Decoder, contains more parameters but demonstrates less tendency to over-fit (Fig \hyperref[fig:1]{1}). In contrast to NAS-based methods\cite{chen2020dip,ho2021neural,arican2022isnas}, our strategy leverages the observed relationship between the architecture and the image to prevent the exhaustive search. Other variants such as DIP-RED \cite{mataev2019deepred} and DIP-TV \cite{liu2019image} augment DIP with additional priors.

\textbf{Early-stopping criterion.} Cheng et al. \cite{cheng2019bayesian} perform posterior inference in DIP so as to prevent the need for early stopping by adding Gaussian noise to the gradients. Similarly, Shi et al. \cite{shi2022measuring} regularize the matrix norm of the network weights to alleviate performance decay. However, tuning the regularization granularity can be challenging (Suppl. B). Jo et al. \cite{jo2021rethinking} combine DIP with Stein's unbiased risk estimator (SURE) \cite{stein1981estimation} for training without clean images, but SURE is limited to only a few known noise types \cite{soltanayev2018training,luisier2010image}. Wang et al. \cite{wang2021early} propose to track the running variance of the outputs, but this also introduces new hyper-parameters that require non-trivial tuning. 


 \section{Investigation and Method}
\subsection{Preliminaries}
\textbf{Deep Image Prior.} A noisy image $\textbf{y}\in\mathbb{R}^N$ can be modeled as: $\textbf{y} = \textbf{x}+\textbf{n}$, 
where $\textbf{x}\in\mathbb{R}^N$ is the clean counterpart to be recovered and \textbf{n} is assumed to be $\mathit{i.i.d.}$ Gaussian Noise drawn from $\mathcal{N}(0,\sigma^2\textbf{I})$ with \textbf{I} being the identity matrix. 
DIP parameterizes the clean image $\mathbf{x}$ via a network $\mathbf{G}_{\theta}$ and is optimized to fit the noisy image $\mathbf{y}$, formulated as:
\begin{equation}
    \theta^\ast = \mathop{\arg \min}\limits_{\theta}\mathcal{L}(\mathbf{y};\mathbf{G_{\theta}(z)}), \quad \mathbf{x}^\ast = \mathbf{G_{\theta^{\ast}}(z)},
\end{equation}
where $\mathbf{z}$ denotes the fixed white noise input tensor. Such parameterization allows lower-frequency contents to be fitted before the higher-frequency ones \cite{chakrabarty2019spectral,shi2022measuring}, exhibiting high impedance to image noises or degradations. Early-stopping is often required to prevent excessive high frequencies from being fitted, such as noise. 

\begin{figure}[t]
\includegraphics[width=0.825\columnwidth]{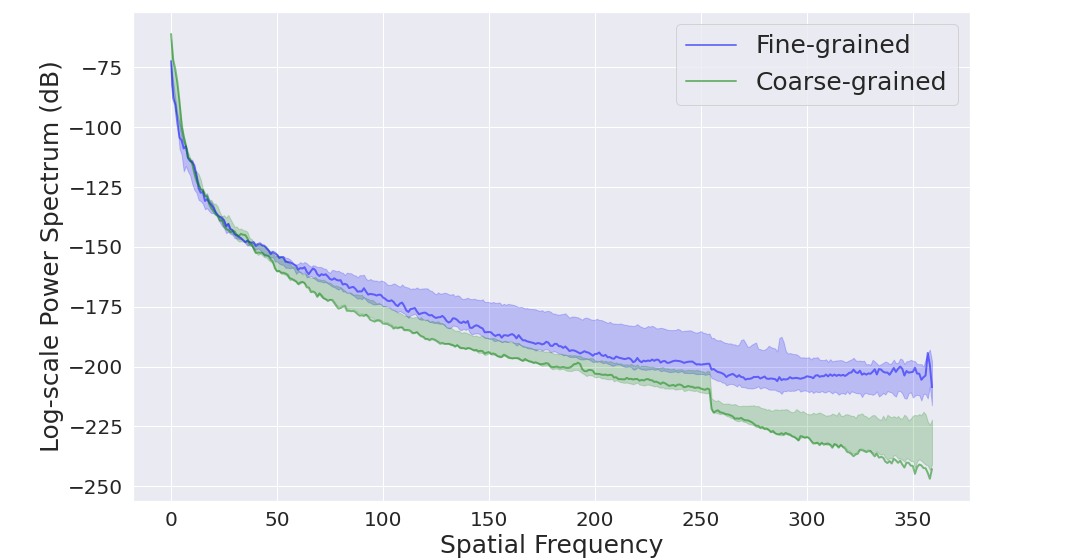}
\label{fig:psd}
\caption{Spectral density of images of different levels of textural complexity, obtained by azimuthally integrateing over the power spectrum of the image.}
\end{figure}

\textbf{Image complexity.} 
Ideally, DIP should denoise an image while preserving textural details. We define image complexity by its texture, which can be characterized by its power spectral density (PSD). The spectral power of a natural image typically follows an exponential decay from low frequencies to high frequencies \cite{simoncelli2001natural}. High-frequency components correspond to fine features such as details, while low-frequency components correspond to coarse structures. Hence, fine-grained images contain more high frequencies than coarse-grained images, giving a flatter PSD curve in (Fig.\hyperref[fig:psd]{2}). Each image is scored based on its texture features, as described in Sec.\hyperref[sec:guidelines]{3.5}.

\subsection{The importance of upsampling} \label{sec:upsampling}
We first identified the core architecture components that affect the denoising performance of DIP. To this end, we analyzed a decoder-only architecture by removing the encoder from the typical encoder-decoder architecture \cite{ulyanov2018deep}, since a decoder is the minimum requirement for reconstructing the final image. Our base model for analysis is a 6-layered convolutional decoder (Conv-Decoder\cite{darestani2021accelerated}), with 128 $3\times 3$ filters per layer except for the last regression layer, followed by batch normalization, ReLU activation function and upsampling. We further simplify the setup by replacing the spatial filters with pixel-wise $1\times 1$ filters, constructing a non-convolutional variant dubbed MLP-Decoder.

\begin{figure} 
\includegraphics[width=\linewidth]{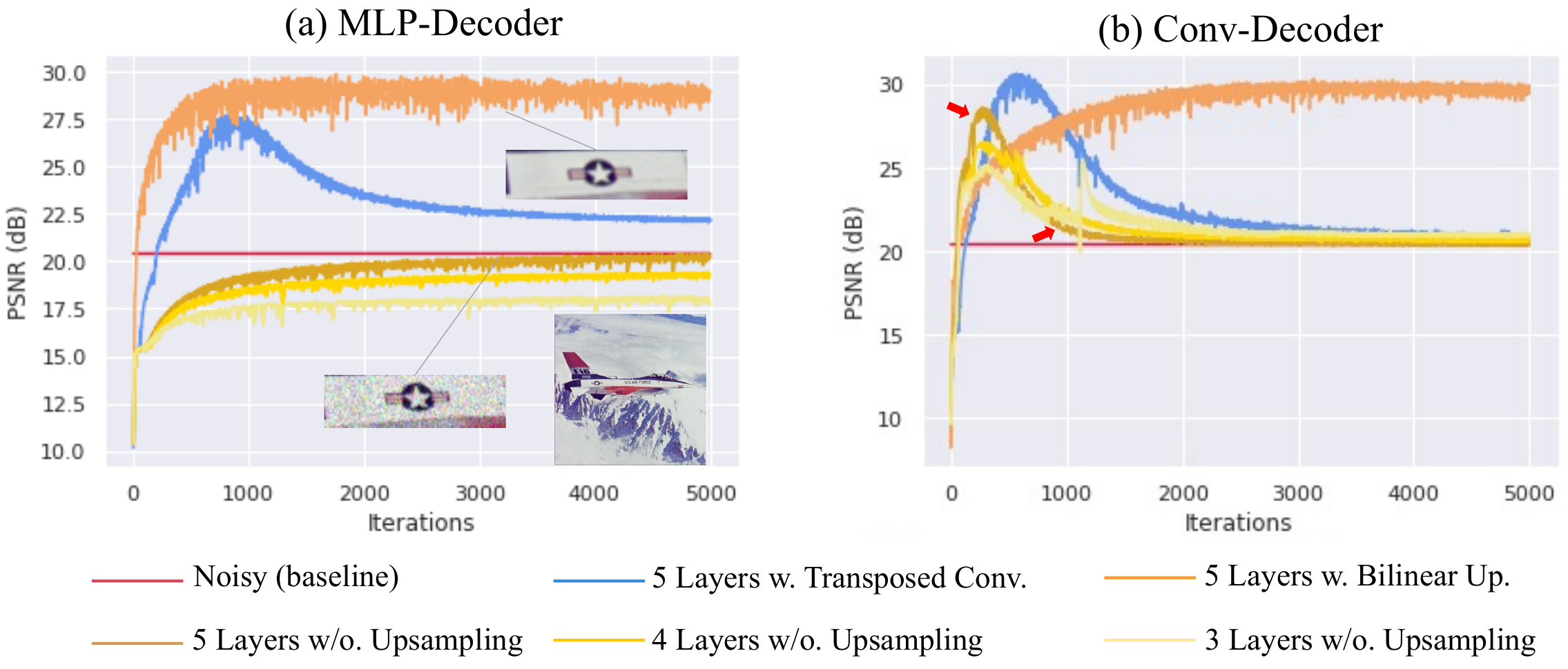}
\caption{\textbf{Influences of architecture components} on image denoising. (\textbf{a}) Without convolutional layers, upsampling still enables denoising. Transposed convolutions lead to faster performance drop due to earlier noise-fitting than bilinear upsampling. Removing the upsampling decreases denoising capability, which cannot be compensated by simply reducing the number of layers, i.e., under-parameterization. (\textbf{b}) Convolutional layers \textit{alone} exhibit certain denoising effects but necessitate early stopping. Increasing the number of convolutional layers achieves higher peak PSNR at the expense of earlier noise fitting (see red arrows). Better results achieved when combined with upsampling.}
\label{fig:upsampling}
\end{figure}




Results for MLP-Decoder, shown in Fig.\hyperref[fig:upsampling]{3} (a), suggest that upsampling plays a vital role. Removal of upsampling results in significant performance degradation of denoising, which cannot be compensated by simply reducing the size of the network for under-parameterization. This holds true for other tasks such as super-resolution (Suppl.A). Fig.\hyperref[fig:upsampling] {3} (b) shows that learnable spatial filters alone also enable denoising, in contrast to the pixel-wise filters, re-affirming the frequency bias of the convolutional layers \cite{chakrabarty2019spectral}. However, the effects vanish as the network size increases and noise-fitting occurs sonner. Convolutional layers together with upsampling achieve better results as manifested in the higher peak PSNR and longer denoising effects. 

\textbf{Discussion.} These results suggest that an appropriate upsampling operation is crucial for effective network image priors, and that under-parameterization alone is insufficient. Different upsampling operations induce varying extents of denoising effects: transposed convolutions \cite{odena2016deconvolution} tend to fit noise faster than bilinear upsampling, necessitating early stopping. In the next section, we investigate the behaviors of these upsampling operations from a signal-processing perspective to gain insights into their influences on denoising performance.

\subsection{Spectral effects of upsampling}
The design choice for upsampling is typically standard: bilinear/nearest neighbor (NN) interpolation or transposed convolutions. These upsampling operations can all be decomposed into two steps: (i) zero-insertion and (ii) filtering. Given a target upsampling factor R, the low-resolution feature map is first interleaved with (R-1) rows/columns of zeros to increase its sampling rate, and then convolved with\begin{center}\begin{minipage}[t]{.63\linewidth} 
\vspace{0pt}
\centering
\includegraphics[width=\linewidth]{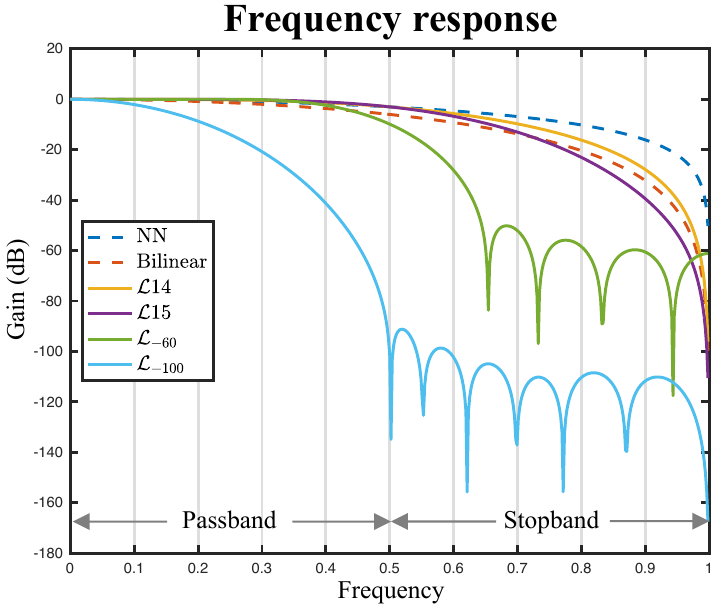} 
\end{minipage}%
\hspace{2pt}
\begin{minipage}[t]{.35\linewidth}
\vspace{0pt}
\begin{adjustbox}{width=\columnwidth}
\begin{tabular}{@{}l|c|ccc}
 LPFs        & Coarse                      & Fine \\ \hline
NN       & 31.1                    & 24.5                     \\ 
Bilinear & 31.3                    & 22.9                       \\ 
$\mathcal{L}14$  & 31.4 & 24.7                       \\
$\mathcal{L}{15}$  & 31.2 & 24.7\\
$\mathcal{L}_{-60}$  & 31.2 & 22.5                       \\ 
$\mathcal{L}_{-100}$ & 27.5 & 20.0                      \\ \hline
NN       & 27.8$\downarrow$                    & \colorbox{lime}{24.2}                       \\ 
Bilinear & \colorbox{lime}{31.1}                     & 22.9                       \\ 
$\mathcal{L}14$  & 25.2$\downarrow$ & \colorbox{lime}{24.0}                       \\ 
$\mathcal{L}{15}$  & 27.1$\downarrow$ & \colorbox{lime}{24.5}                       \\ 
$\mathcal{L}_{-60}$  & \colorbox{lime}{31.2} & 22.5                       \\ 
$\mathcal{L}_{-100}$ & 27.5 & 20.0                       \\

\end{tabular}
\end{adjustbox}
\end{minipage} \label{fig:responses}
\captionof{figure}{(\textbf{Left}) Frequency responses of the tested LPFs. Different LPFs result in upsamplers with different extents of smoothing. NN interpolation preserves most signals in the passband but also the high-frequency replica in the stopband; $\mathcal{L}_{-100}$ attenuates the signals most significantly ($\sim100dB$) and suppresses the high-frequency replica most. (\textbf{Right}) Denoising results on coarse- and fine-textured images from Set9. \textbf{Top rows}: Peak PSNR values. \textbf{Bottom rows}: PSNR values at the last training iteration. }
\end{center} a low-pass filter (LPF) to remove the alias high frequencies introduced by zero-insertion. The key difference between the upsampling operations lies in the nature of the filters. The filters for transposed convolutions are learnable, but are fixed for bilinear and nearest nighbor upsampling.

To better illustrate this, we consider the case of a 1D signal $x(n), n=0,...,N-1$, and its discrete Fourier representation $X(k)=\sum_{n=0}^{N-1}x(n)e^{\frac{-i2\pi}{N}kn}, k=0,...,N-1$. 

For an upsampling factor of 2, we have
\begin{align}
&X^{\text{up}}(\hat{k}) = \sum_{n=0}^{2N-1} x^{\text{up}}(n)e^{\frac{-i2\pi}{2N}\hat{k}n} \\
                    &= \sum_{n=0}^{N-1} x^{\text{up}}(2n) e^{\frac{-i2\pi}{2N}(2n)\hat{k}} + \sum_{n=0}^{N-1} x^{\text{up}}(2n+1) e^{\frac{-i2\pi}{2N}(2n+1)\hat{k}}, 
\end{align}
where $\hat{k}=0,\hdots,2N-1$, $x^{\text{up}}(2n)=x(n)$, $x^{\text{up}}(2n+1)=0$ due to interleaved zero insertion. Hence, for $0\leq\hat{k}<N$, $X^{\text{up}}(\hat{k})=X(k)$. 

For $\hat{k}\geq N$, let $k'=\hat{k}-N$, $k'=0,...,N-1$, we have
\begin{align}
    X^{\text{up}}(\hat{k}) &= \sum^{N-1}_{n=0} x(n)e^{\frac{-i2\pi}{2N}(k'+N)2n} \\
                &= \sum^{N-1}_{n=0} x(n)e^{\frac{-i2\pi}{N}nk'-i2n\pi} =X(k')\label{eq:5}, 
\end{align}  where Eq.\ref{eq:5} exploits the periodicity of the complex exponential function. 
Thus, zero-insertion will preserve the original spectrum at $[0,N)$ (passband) and additionally create a mirrored copy at $[N,2N-1]$ (stopband). The high-frequency replica beyond $N$ should be suppressed by the subsequent LPF to avoid image artifacts. According to duality and convolution theorem, convolving with NN or bilinear interpolation filter is equivalent to multiplication of $X^{up}(\hat{k})$ with a \texttt{Sinc} or \texttt{Sinc}$^2$ function corresponding to low-pass filtering, while transposed convolutional filter may not necessarily be low-pass as it depends on the optimization objective. 

\begin{table}[t] \label{ta:iteration}
\small
\begin{adjustbox}{width=0.8\columnwidth}
\begin{tabular}{@{}clcccc@{}}
\toprule
                     & \multicolumn{1}{c}{NN}   & $\mathcal{L}$14            & $\mathcal{L}$15           & $\mathcal{L}$16                & $\mathcal{L}$17               \\ \midrule
Coarse-grained       & 1783                     & 1589             & 1681            & 2205                 & 2214                 \\
Fine-grained         & \multicolumn{1}{c}{2424} & 2942             & 3898            & 4361                 & 4957                 \\ \bottomrule
\end{tabular}
\caption{\label{ta:iter_num}The iteration number [iter./5000] at the peak PSNR for different upsamplers. The upsamplers are designed to mainly differ in the stopband. The peak PSNR is reached more slowly when the attenuation is stronger (from left to right).}
\end{adjustbox}
\end{table}


We experimentally demonstrate that different upsampling operations bias the architecture towards different spectral properties. Specifically, we constructed four upsamplers by first interleaving the input with zeros and then convolving it with handcrafted LPFs (shortened as $\mathcal{L}$): $\mathcal{L}14$, $\mathcal{L}$15, $\mathcal{L}_{-60}$ and $\mathcal{L}_{-100}$, with the subscript denoting the decayed dB. By construction, $\mathcal{L}{14}$ and $\mathcal{L}{15}$ are very close to NN in the passband ($<0.03dB$) and only differ in the stopband. The frequency responses of the compared LPFs are detailed in Fig. \hyperref[fig:responses]{4}. We applied the customized upsamplers on ConvDecoder and tested them on the fine- and coarse-textured images from Set9 respectively. Similar findings hold for the encoder-decoder architecture (Suppl. C).

From Fig. \hyperref[fig:responses]{4}, upsampling critically influences both the peak PSNR value and the timing for early stopping with respect to images of different textural complexities. Upsamplers (NN, $\mathcal{L}14$, $\mathcal{L}{15}$) with less attenuation on the high-frequency replica are beneficial for generating fine-grained images, but they tend to cause noise-fitting especially for coarser-grained images. Also, they are generally the fastest to reach the peak PSNR (Table \hyperref[ta:iteration]{1}). This explains why the transposed convolution requires early-stopping, as the filters may not learn to attenuate the introduced high frequencies effectively. Similar issues with the learned upsampling are also prevalent in generative models \cite{schwarz2021frequency,chandrasegaran2021closer,durall2020watch,frank2020leveraging}. 

On the other hand, bilinear and $\mathcal{L}_{-60}$ exhibit a greater amount of attenuation, more strongly biasing the network against high frequencies and leading to longer-lasting denoising effects. They turn out to work sufficiently well for both kinds of images, especially on the coarser-grained ones which are typically the majority in the dataset. 
LPF$_{-100}$ tends to over-smooth the output and performs the worst as it attenuates both the passband and stopband signals substantially, though not requiring early stopping. 

\textbf{Discussion.} These results lead us to conclude that the upsamplers with fixed LPFs are key to the denoising effects of DIP for their tendency towards smooth images (i.e., reduced high-frequency contents), which aligns well with the spectral statistics of natural images (Fig.\hyperref[fig:psd] {2}). Probably due to a good balance between the denoising performance and persistency, bilinear upsampling has been widely adopted in DIP models for various applications \cite{ulyanov2018deep,heckel2018deep,gandelsman2019double, heckel2019denoising}.

\subsection{Interactions with other architecture elements}
After establishing the significance of upsampling, we will now consider how it may interact with other common architectural elements in affecting the ultimate output.

\begin{figure}
\includegraphics[width=\linewidth]{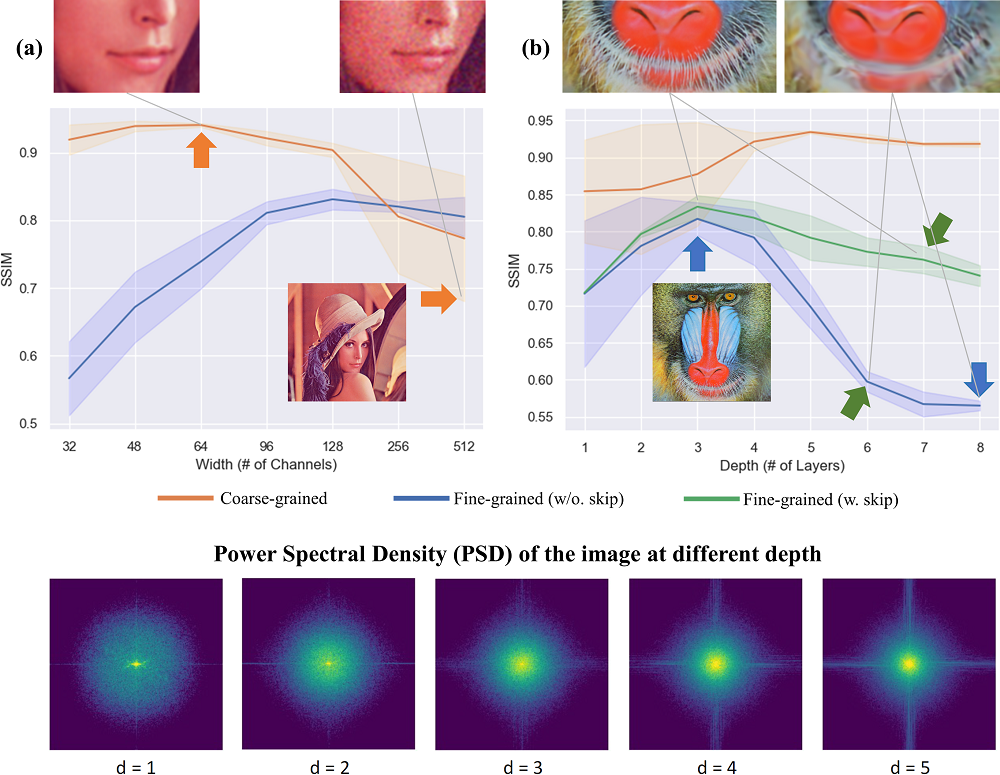} \label{fig:elements}
\caption{Influences of width, depth, and skip connections, assuming upsampling inserted in-between the layers. \textbf{(a)} Tendency to over-fit coarse-grained images increases with width. SSIM scores averaged across the depths. \textbf{(b)} Tendency to over-smooth fine-grained images increases with depth, but alleviated by skip connections. SSIM scores are averaged across the widths.} 
\end{figure}

\textbf{Convolutions + non-linearity}. Ideal upsampling does not modify the signal representations but only expands the spectrum for the subsequent layers to add new frequency contents. Convolution followed by nonlinearity such as ReLU, is the only operation capable of introducing arbitrarily high frequencies \cite{karras2021alias}. Increasing the number of layers (depth) or channels (width) enhances the capability of generating new high frequencies, as theoretically and empirically proved in \cite{rahaman2019spectral}.

\begin{figure}
\includegraphics[width=0.8\linewidth]{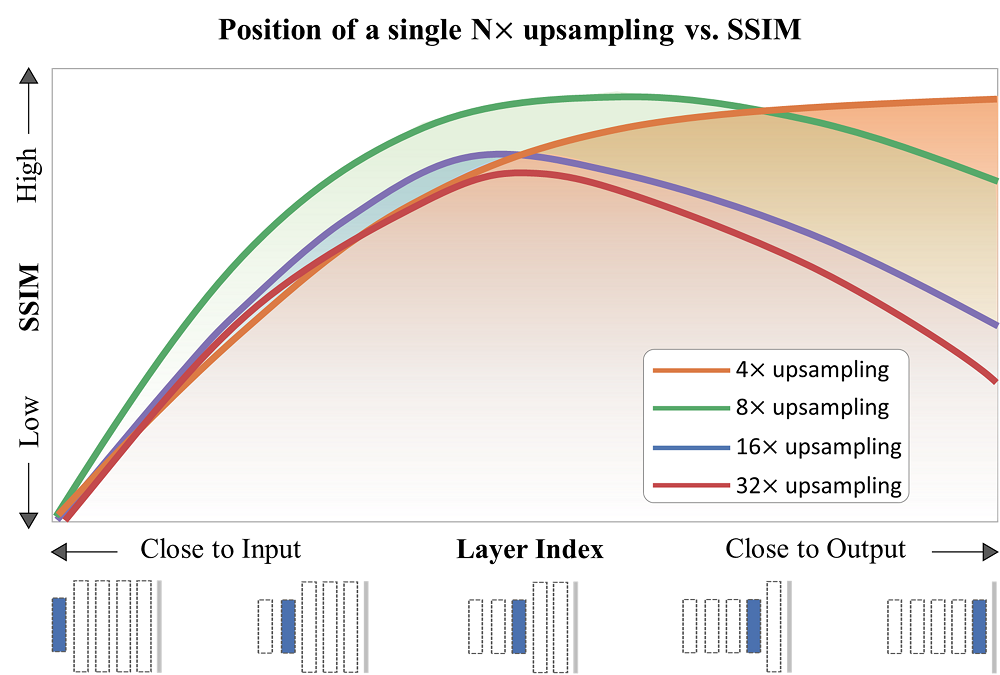} \label{fig:upsamplingPos}
\caption{Position of upsampling. Placing the upsampling close to the input (or encoder) can easily cause over-fitting, regardless of the scaling factor. When close to the output (end of the decoder), upsampling with large scaling factors (e.g., 32$\times$)  causes over-smoothing. }
\end{figure}

Intuitively, using an excessive number of layers or channels can accelerate the learning of both details and noise. However, the effects can be attenuated by the fixed upsampling operations between the layers. As shown in  Fig.\hyperref[fig:elements] {5}, when using fewer upsampling operations (i.e., a shallower network), increasing the width of the network increases the tendency to cause overfitting on simpler images while benefiting the more complex images. Increasing the number of upsampling operations (i.e., a deeper network) can alleviate over-fitting with stronger attenuation but results in blurry outputs for fine-grained images. Increasing only the upsampling factors without adding more layers can make the output even more blurry (Fig.\hyperref[fig:upsamplingPos] {6}). In other words, the final output is determined by the balance between the generation of high frequencies by the layers and signal attenuation caused by the upsampling operations. 



\textbf{Skip connections} between the encoder and the decoder often complicate the design space \cite{chen2020dip}. While they are not directly responsible for denoising, they may lower the effective upsampling rate, making deep networks perform similarly to shallower ones, i.e., resulting in lesser suppression of the high-frequency replica. This finding is validated on a large-scale experiment comprising $7424$ architectures (Suppl. D). As examplified in Fig. \hyperref[fig:elements] {5} (b), skip connections notably ameliorate the over-smoothing issue induced by the same deep network. Overall, skip connections exert a more pronounced influence on the deeper networks compared to the shallower ones, as evident in the smaller deviation observed when depth $\leq$3 in Fig.\hyperref[fig:elements] {5} (b). 





\begin{table*}[hbt]
\caption{\textbf{Quantitative results on Gaussian noise}. $\sigma$ denotes the noise level. All methods were trained with \textbf{a fixed iteration number (3000)} throughout the experiments. The highest score is in \textbf{bold}, and the second highest is \underline{underlined}. } \label{ta:gauss}
\small
\centering
\renewcommand{\arraystretch}{1.0}
\begin{adjustbox}{width=0.85\columnwidth,center}
\begin{tabular}{@{}cccccccl@{}}
\toprule
Datasets                &                         & \multicolumn{1}{c}{DIP \cite{ulyanov2018deep}}   & \multicolumn{1}{c}{Deep Decoder \cite{heckel2018deep}} & \multicolumn{1}{c}{ConvDecoder \cite{darestani2021accelerated}} & \multicolumn{1}{c}{NAS-DIP \cite{chen2020dip}} & \multicolumn{1}{c}{ISNAS-DIP \cite{arican2022isnas}} & \multicolumn{1}{c}{Ours}                              \\ \midrule 
& & & &   \multicolumn{2}{c}{$\sigma=25$} \\ \cmidrule(lr){3-8}
\multirow{2}{*}{Set9 \cite{dabov2007video}} & \multicolumn{1}{c}{PSNR}                     & \underline{30.10} & 28.45                            & 28.51                           & 26.37   & 29.11                       & \multicolumn{1}{c}{\textbf{30.26}}   \\
                      & \multicolumn{1}{c}{SSIM}                     & \underline{0.893} & 0.848                            & 0.854                           & 0.753   & 0.862                        & \multicolumn{1}{c}{\textbf{0.900}}  \\ 
\multirow{2}{*}{Set12 \cite{zhang2017beyond}}                     & \multicolumn{1}{c}{PSNR}                     & \underline{26.97} & 25.98                        & 25.78                            & 20.86   & 24.10                    &    \multicolumn{1}{c}{\textbf{28.14}}                                   \\
                                           & \multicolumn{1}{c}{SSIM} & \underline{0.812} & 0.789                           & 0.786                          & 0.534   & 0.745                         &    \multicolumn{1}{c}{\textbf{0.884}}                                 \\ 
\multirow{2}{*}{CBSD68 \cite{roth2009fields}}                     & \multicolumn{1}{c}{PSNR}                     & \textbf{28.93} & 25.50                            & 25.19                           & 23.80   & 24.51                         &  \multicolumn{1}{c}{\underline{28.57}}                   \\
                                 &  \multicolumn{1}{c}{SSIM}                     & \textbf{0.892} & 0.809                            & 0.793                           & 0.693   &   0.745                            & \multicolumn{1}{c}{\underline{0.888}}                                     \\ \midrule  & & & &  \multicolumn{2}{c}{$\sigma=50$}\\ \cmidrule(lr){3-8}
\multirow{2}{*}{Set9 \cite{dabov2007video}} & \multicolumn{1}{c}{PSNR}                   & 25.04 &                         \underline{25.22}   &  25.01           &    21.07                      & 23.91      & \multicolumn{1}{c}{\textbf{26.13}}\\
                       & \multicolumn{1}{c}{SSIM}                     & 0.761  & 0.764                           & \underline{0.769}                           & 0.593   &  0.698                        & \multicolumn{1}{c}{\textbf{0.833}} \\ 
\multirow{2}{*}{Set12 \cite{zhang2017beyond}}                     & \multicolumn{1}{c}{PSNR}                     & 22.15  & 20.44                           & \underline{22.72}                          & 18.92   & 19.20                          &              \multicolumn{1}{c}{\textbf{24.59}}                      \\
                                    & \multicolumn{1}{c}{SSIM} & 0.623  & 0.687                   & \underline{0.706}                           & 0.476   & 0.537        & \multicolumn{1}{c}{\textbf{0.805}}                                 \\ 
\multirow{2}{*}{CBSD68 \cite{roth2009fields}}                     & \multicolumn{1}{c}{PSNR}                     & 23.74  & 23.52                            & \underline{24.06}                          & 17.92    & 19.93                         &  \multicolumn{1}{c}{\textbf{24.17}}                                  \\
                                          & \multicolumn{1}{c}{SSIM}                     & 0.746  & 0.725                            & \underline{0.767}                           & 0.323   &    0.573                           &  \multicolumn{1}{c}{\textbf{0.774}}                                  \\ \midrule  \rowcolor[gray]{0.9} \multicolumn{1}{l}{\# Params (Millions)} &  & 2.3M & \textbf{0.1M} & 0.89M &  4.4M & \emph{Varied} & \multicolumn{1}{c}{\textbf{0.05M$\sim$0.92M}} \\ \bottomrule 
\end{tabular}
\end{adjustbox}
\end{table*} 

\begin{table}
\begin{adjustbox}{width=0.86\columnwidth}
\begin{tabular}{@{}|c|c|c|c|@{}} 
 \hline
 & NAS \cite{chen2020dip} & ISNAS \cite{arican2022isnas} & Ours \\ [0.5ex] 
 \hline\hline
 Image-Specific  &     \color{BrickRed}{\xmark}                     &    \color{ForestGreen}\checkmark                        & \color{ForestGreen}{\checkmark}                          \\ \hline
Architecture Search        &    3 days                      &    5 mins                        &    --                       \\ \hline
\textit{Per-image} Restoration             &  $\sim$23 mins                        &  $\sim$7 hrs                          & $\sim$6 mins                          \\ \hline
Early Stopping Required? &  Yes                        &   Yes                         &   \color{ForestGreen}No \\ 
 \hline
\end{tabular}
\caption{\textbf{Comparisons of desired properties.} Restoration time is computed on an image of size $512\times512$ with 3000 iterations.} \label{ta:nas-comp}
\end{adjustbox}
\end{table}


\subsection{Practical application to architectural design} \label{sec:guidelines}
Based on the above findings and analysis, we argue that it is possible to estimate an effective architecture for each image without extensive search. 
Assuming every decoder layer is followed by a $2\times$ bilinear upsampling layer, our strategies are as follows:

\textbf{Depth estimation}. We have shown that increased depth tends to over-smooth the output, affecting fine-grained images much more than on coarse-grained images. For fine-grained images, we have three options: \textbf{a}) add more skip connections to a deep network; \textbf{b}) simply use a shallower one; \textbf{c}) keep all the layers but reduce down-/up-sampling layers. We recommend \textbf{c} for decoder-only architectures since they are already under-parameterized; for hourglass network this can easily lead to over-fitting (Fig.\hyperref[fig:upsamplingPos] {6}). We find \textbf{b} is generally better than \textbf{a} in trading off between good performance and the need for early stopping, especially under higher-level noise, as shown in our results section. This also holds for coarse-grained images as they are not sensitive to depth. More specifically, we find a 2-level hourglass network sufficient for both types of images (Suppl. D).
    
\textbf{Width estimation}. Width is crucial for learning sufficient details while avoiding over-fitting especially to a shallow network (less signal attenuation). We find that the width needs to be set according to textural complexity of the image: a finer-grained image requires more channels per layer and vice versa. To further validate this, we treated width estimation as a classification problem, and trained three SVMs \cite{cortes1995support} with texture features as the inputs to classify the images into three width choices $\{32, 64, 128\}$. Note that these widths were empirically chosen for the datasets used in this work and are by no means optimal for all cases, but tuning for other images should be straightforward. The classification results and analysis are in Sec.\hyperref[sec:more]{4.4}.  

\begin{table}
\begin{center}
\begin{adjustbox}{width=\columnwidth,center}
\small
\begin{tabular}{@{}clllllll@{}} 
\toprule
Noise scale                              &      & \multicolumn{1}{c}{DIP}            & \multicolumn{1}{c}{DD}    & \multicolumn{1}{c}{CD} & \multicolumn{1}{c}{NAS} & \multicolumn{1}{c}{ISNAS} & \multicolumn{1}{c}{Ours} \\ \midrule
\multirow{2}{*}{$\zeta=0.01$}                    & PSNR & \textbf{31.58} & 29.81                     & 29.51                  & 29.70                   &  30.12                         & {\underline{30.95}}              \\
                                         & SSIM & \textbf{0.915} & \multicolumn{1}{c}{0.880} & 0.875                  & 0.864                   & 0.875                          & \underline{0.910}             \\
\multirow{2}{*}{$\zeta=0.1$}                     & PSNR & 22.66          & 23.91                     & \underline{24.94}            & 15.72                  &  17.68                         & \textbf{24.99}           \\
                                         & SSIM & 0.640          & 0.718                     & \underline{0.776}            & 0.417                   &  0.496                         & \textbf{0.789}           \\
\multirow{2}{*}{$\zeta=0.2$} & PSNR & 20.36         & 21.13                     & \underline{21.87}            & 12.99                   &  14.64                         & \textbf{22.55}           \\
\multicolumn{1}{l}{}                     & SSIM & 0.563          & 0.609                     & \underline{0.672}           & 0.311                   &  0.383                         & \textbf{0.774}           \\ \bottomrule
\end{tabular}
\caption{\textbf{Quantitative evaluation on Poisson noise}. Deep Decoder shortened as DD and ConvDecoder as CD.} \label{ta:pos-noise}
\end{adjustbox}
\end{center}
\end{table}

\textbf{Image scoring}. To more robustly classify image texture, we extracted both spatial and frequency features and trained a Decision Tree \cite{loh2011classification} for feature selection. We used the classic Gray Level Co-occurrence Matrix (GLCM) \cite{haralick1973textural} for deriving the spatial texture features from each image. The most useful features turn out to be the following: correlations measured at $0^{\circ}$, homogeneity at $45^{\circ}$, and contrast at $0^{\circ}$. Frequency features are captured using a 1D PSD vector obtained by first converting the 2D PSD from Cartesian coordinates $S(k,l)$ to polar coordinates, and then azimuthally averaging over $\theta$, defined as $\mathit{\hat S}(r) = \frac{1}{2\pi}\int^{2\pi}_{0}\mathit{S}(r,\theta)d\theta$ with $r=\sqrt{k^2+l^2}$ and $\theta=\atantwo(k,l)$, which represents the mean magnitude of the frequencies with respect to the radial distance $r$. Refer to Suppl. E for more details.

\begin{figure*}[ht]
\includegraphics[width=0.75\linewidth]{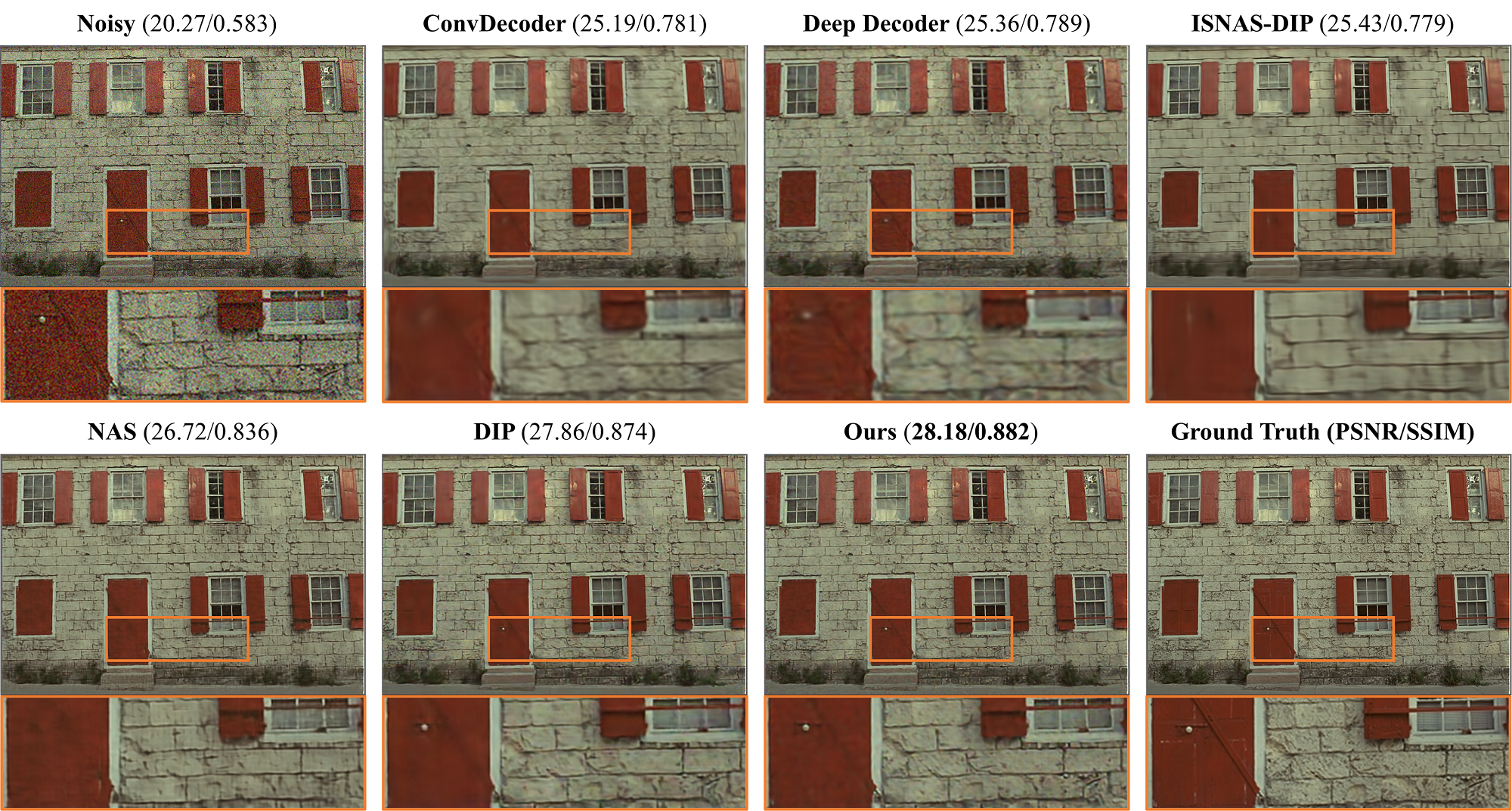} \label{fig:gaussian}
\caption{Denoising results on a \textbf{fine-grained} image ("kodim01" from Set9) with \textbf{Gaussian noise} ($\sigma=25$). Our estimated architecture for this image is a two-level hourglass network with one skip connection and 128 channels, which is much smaller than others.}
\end{figure*}

\section{Experiments}

\subsection{Implementation Details}
We conducted the experiments on three popular datasets and a real-world noisy dataset: \textbf{Set9} \cite{dabov2007video} consisting of 9 colored images, \textbf{Set12} \cite{zhang2017beyond} consisting of 12 grey-scaled images, \textbf{CBSD68} \cite{roth2009fields} consisting of 68 colored images, and \textbf{PolyU} \cite{xu2018real} consisting of 100 real noisy and clean image pairs. We first report our results with \textbf{2-level} hourglass architecture with the same components as in DIP \cite{ulyanov2018deep} when comparing with the existing methods, and then extend our strategy to ConvDecoder \cite{darestani2021accelerated}, a decoder-only architecture. All models were trained for 3000 iterations.

\subsection{DIP variants}

\begin{figure}
\includegraphics[width=0.9\linewidth]{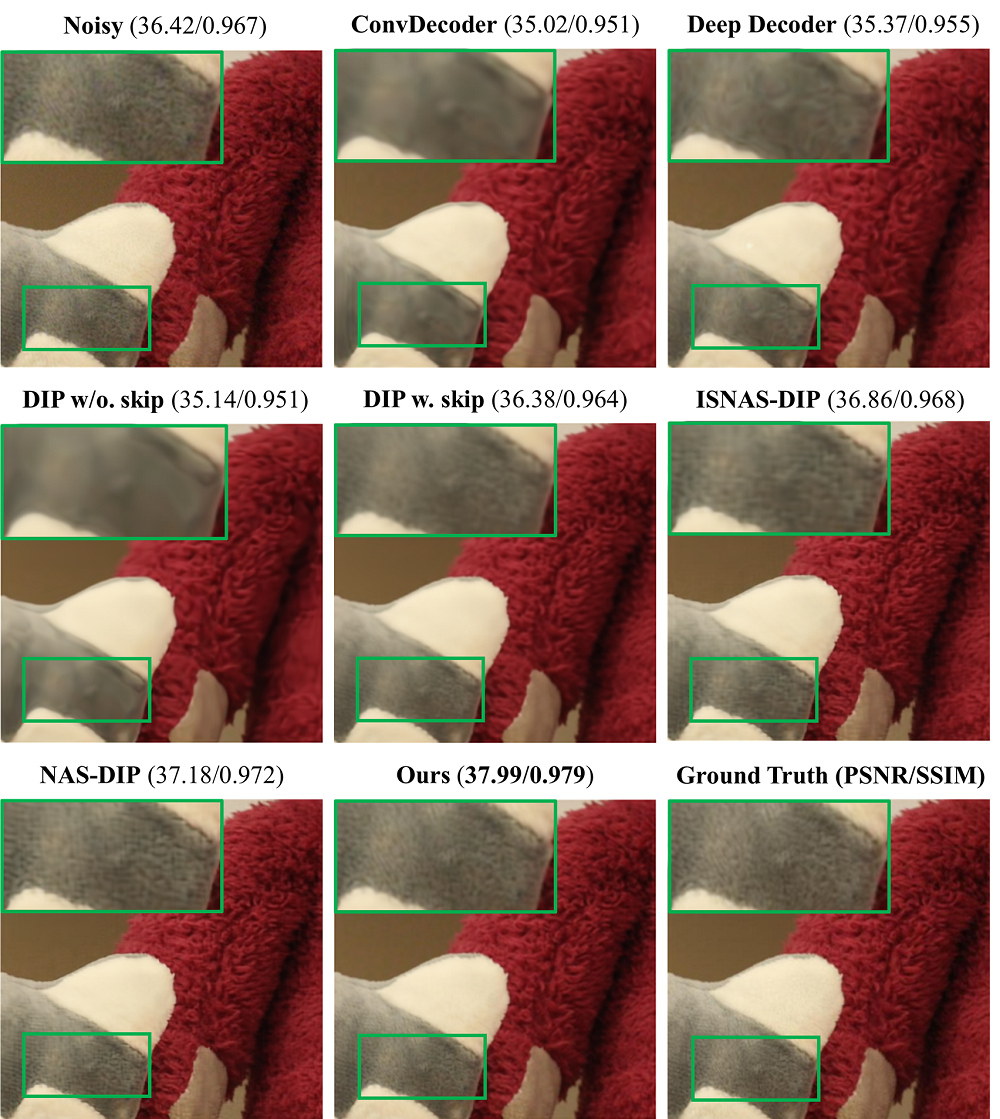} \label{fig:real_world}
\caption{Denoising results on a \textbf{real-world noisy} image.} \label{fig:realnoise}
\end{figure}

\textbf{Gaussian Noise.} Compared with the base DIP network, the properly-designed under-parameterized networks estimated with our strategy perform on par at a mild noise level while excel at a higher noise level (Table \hyperref[ta:gauss] {2}). This cannot be solely explained by under-parameterization since Deep Decoder and ConvDecoder contain similar or even fewer parameters while are unable to achieve similar results. Fig. \hyperref[fig:gaussian] {7} shows that a shallow and wide network can preserve the details better than many deeper ones. Note that DIP is a 5-level hourglass network with full skip connections, which by our standard can also well preserve the details. In fact, it is a very strong baseline under mild noise. NAS-DIP and ISNAS-DIP suffer from various degrees of over-fitting on different datasets. This also suggests that the point of optimal stopping varies across images. Besides, they are time-intensive in either searching or evaluation (Table \hyperref[ta:nas-comp] {3}).
Similar conclusions hold for \textbf{Poisson noise}, which was tested on Set9 \cite{dabov2007video} as shown in Table \hyperref[ta:pos-noise] {4}. 

\begin{table}[ht]
\begin{center}
\begin{adjustbox}{width=0.95\columnwidth}
\small
\begin{tabular}{@{}lllllll@{}}
\toprule
    & \multicolumn{1}{c}{DIP} & \multicolumn{1}{c}{DD} & \multicolumn{1}{c}{CD} & \multicolumn{1}{c}{NAS} & \multicolumn{1}{c}{ISNAS} & Ours  \\ 
    \midrule
   PSNR & \textbf{38.15}                   & 37.22                  & 37.00                  & 37.83                     &    37.78                       &  38.05 \\
   SSIM & 0.982                   & 0.978                  & 0.976                  & 0.982                        &      0.977                    & \textbf{0.984} \\ 
\bottomrule
\end{tabular}
\caption{\textbf{Quantitative evaluation on PolyU, a real-world noisy dataset}. Deep Decoder and ConvDecoder shortened as DD, CD.} \label{ta:realNoise}
\end{adjustbox}
\end{center}
\end{table}

\textbf{Real-World Noise.} We additionally evaluated all methods on the PolyU dataset \cite{xu2018real}. We applied again the two-level hourglass architectures with variable width estimated for the images. Table \hyperref[ta:realNoise] {5} summarizes the numerical results. Fig.\hyperref[fig:realnoise] {9} shows that our method tends to preserve more details, though this may not be reflected by the metrics. Simply removing the skip connections from DIP causing blurring, similar to the decoder-only architectures. 

\begin{table}[h]
\captionsetup{font=small}
\caption{\textbf{Application to ConvDecoder}.} \label{ta:decoders}
\begin{adjustbox}{width=\columnwidth,center}
\begin{tabular}{@{}cccccc@{}}
\toprule
\multirow{2}*{Datasets}       &  & \multicolumn{2}{c}{$\sigma=25$} &   \multicolumn{2}{c}{$\sigma=50$}   \\ \cmidrule(lr){3-4} \cmidrule(lr){5-6} & &  Before &   \colorbox[gray]{0.9}{After}  & Before & \colorbox[gray]{0.9}{After}  \\ 
\midrule 
\multirow{2}*{Set9} & \multicolumn{1}{c}{PSNR} & 28.51   & \colorbox[gray]{0.9}{28.74}  & 25.01 & \colorbox[gray]{0.9}{25.11} \\ 
&  \multicolumn{1}{c}{SSIM } & 0.854  & \colorbox[gray]{0.9}{\textbf{0.873}}  & 0.769 & \colorbox[gray]{0.9}{\textbf{0.784}} \\
\multirow{2}*{Set12} & \multicolumn{1}{c}{PSNR} & 25.79   & \colorbox[gray]{0.9}{\textbf{26.98}} & 22.72 & \colorbox[gray]{0.9}{23.01} \\ 
& \multicolumn{1}{c}{SSIM } &  0.786  & \colorbox[gray]{0.9}{\textbf{0.854}}  & 0.706 & \colorbox[gray]{0.9}{\textbf{0.742}} \\
\multirow{2}*{CBSD68} & \multicolumn{1}{c}{PSNR} & 25.19   & \colorbox[gray]{0.9}{\textbf{28.29}}  & 24.36 & \colorbox[gray]{0.9}{24.12} \\ 
& \multicolumn{1}{c}{SSIM } &  0.793  & \colorbox[gray]{0.9}{\textbf{0.877}}  & 0.767  & \colorbox[gray]{0.9}{0.768} \\ 
\midrule   
\multicolumn{1}{l}{\# Params (Millions)} &  & 0.89M & \colorbox[gray]{0.9}{0.06M$\sim$0.89M} & 0.89M & \colorbox[gray]{0.9}{0.06M$\sim$0.89M} \\
\bottomrule
\end{tabular}
\end{adjustbox}
\end{table}

\begin{figure}[h]
\includegraphics[width=\linewidth]{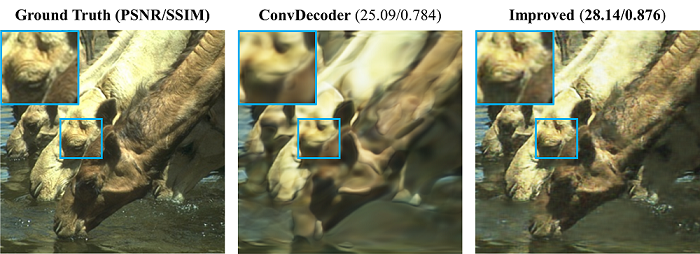} \label{fig:convdecoder}
\caption{\textbf{Qualitative improvement on ConvDecoder} simply by removing three upsampling layers in this case while preserving all the convolutional layers.}
\end{figure}

\subsection{Decoder-only Architectures}
We applied our strategy to ConvDecoder and tested it on all three datasets. Here we keep all 5 layers but remove a certain number of upsampling layers and scale the width accordingly for the images. These simple changes effectively alleviate the over-smoothing issue that often brought by ConvDecoder as shown in Fig.\hyperref[fig:convdecoder] {8} and improve the quantitative results as shown in Table \hyperref[ta:decoders] {6}.

\subsection{More Analysis on Depth and Width} \label{sec:more}
\begin{figure}
\includegraphics[width=\linewidth]{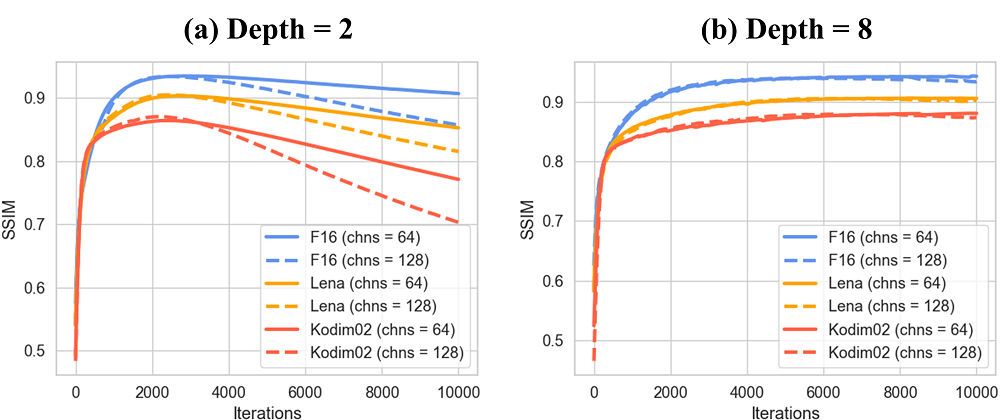} \label{fig:shallow-deep}
\caption{\textbf{(a)} Width critically influences a shallower network, while \textbf{(b)} it rarely has any impact on a sufficiently deep network.}
\end{figure}

\begin{figure}
\includegraphics[width=0.9\linewidth]{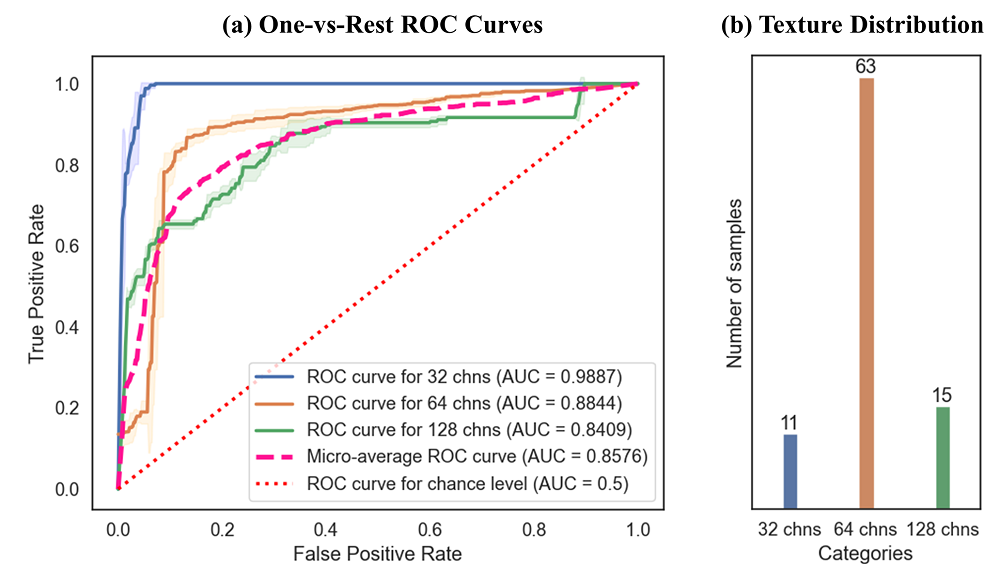} \label{fig:rocCurve}
\caption{\textbf{(a) ROC curves with AUC scores for width classifcation} on images from Set9, Set12 and CBSD68. 5-fold cross-validation is performed and repeated 10 times. \textbf{(b)} Overview of \textbf{our Texture-DIP Dataset}.}
\end{figure}

 In practice, we find it more efficient to first determine the depth, and then the width. This is because when the network is deep enough (strong attenuation due to upsampling),  width becomes less influential, as evident in Fig.\hyperref[fig:shallow-deep] {10} (b) and the small deviation in Fig.\hyperref[fig:elements] {5} (b). However, to relax the need for early stopping at a higher noise level, one may prefer a shallower one, where width matters more (Fig.\hyperref[fig:shallow-deep] {10} (a)). In this regard, we use the texture features of the images to predict the desired width. This makes intuitive sense as noise-fitting is associated with the amount of high frequency contents in the image, which is manifested in its texture. This is corroborated by the 0.86 Micro-average AUC score shown in Fig.\hyperref[fig:rocCurve] {11} (a). The "optimal" width labels we used to train the classifiers were obtained by experimenting with all three width choices on a 2-level hourglass architecture. These width labels are also applicable to decoder architectures as demonstrated by our experiments. 
 
 Although the width choices seem limited, we found the images robust to the choice of width to some extent, and simple averaging also works well for ambiguous cases (Appendix D). In fact, some images have multiple width labels. We included these cases in the released Texture-DIP dataset. 

\subsection{What about Transformers?}
\begin{figure*}
\includegraphics[width=0.8\columnwidth]{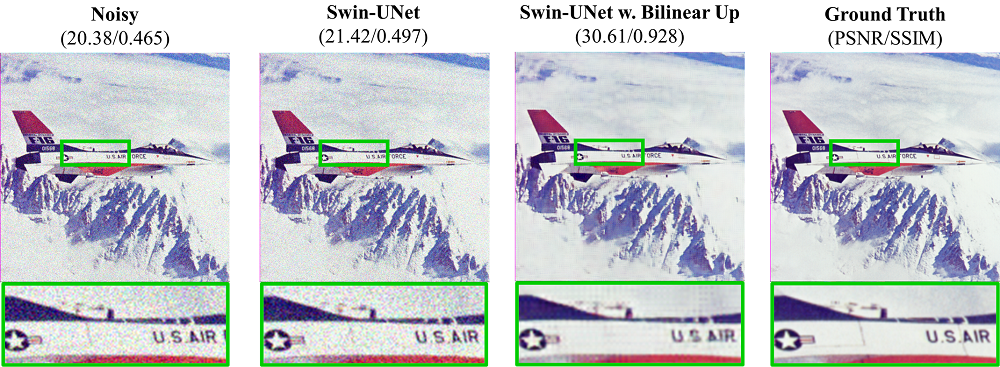} \label{fig:transformer}
\caption{Visual comparisons on the transformer-based Swin-UNet \cite{cao2022swin} with and without bilinear upsampling. }
\end{figure*}

Transformers \cite{vaswani2017attention} have become integral parts of modern deep neural network architectures. We experimented with Swin-UNet \cite{cao2022swin}, an encoder-decoder pure transformer consisting of Swin Transformer blocks \cite{liu2021swin} and skip connections. Note that Swin-UNet relies on patch merging/expanding layers for $2\times$ down-/upsampling without involving unlearnt upsampling. We replaced all the patch expanding layers with $2\times$ bilinear upsampling operations and compared the performance of the modified network with the original. We did not replace the patch merging layers with the corresponding downsampling layers or strided convolutions because we did not observe any significant difference. The results presented in Fig. \hyperref[fig:transformer]{12} are consistent with our findings on CNNs. For more implementation details, please refer to our code.

\section{Conclusion and Future Work}
We presented simple but effective solutions to the challenging DIP architectural design in the context of image denoising. Leveraging the spectral effects of upsampling and its relationships with other architectural compoenents, we show that simple architectural changes yield highly-effective under-parameterized networks that could surpass the larger counterparts and does not critically rely on early-stopping. Although this work focuses on denoising, insights learned about upsampling can be employed in the future to understand the relationship between architectural characteristics and other image restoration tasks. We hope our study could encourage efficient architectural design for DIP and image synthesis in general. 

\section*{Acknowledgements}
This work was supported in part by the United States National Institutes of Health (NIH) under Grant CA266702 and Grant EB008374.


{\small
\bibliographystyle{ieee_fullname}
\bibliography{egbib}

\begin{thebibliography}{10}\itemsep=-1pt

\bibitem{arican2022isnas}
Metin~Ersin Arican, Ozgur Kara, Gustav Bredell, and Ender Konukoglu.
\newblock Isnas-dip: Image-specific neural architecture search for deep image
  prior.
\newblock In {\em Proceedings of the IEEE/CVF Conference on Computer Vision and
  Pattern Recognition}, pages 1960--1968, 2022.

\bibitem{cao2022swin}
Hu Cao, Yueyue Wang, Joy Chen, Dongsheng Jiang, Xiaopeng Zhang, Qi Tian, and
  Manning Wang.
\newblock Swin-unet: Unet-like pure transformer for medical image segmentation.
\newblock In {\em European conference on computer vision}, pages 205--218.
  Springer, 2022.

\bibitem{chakrabarty2019spectral}
Prithvijit Chakrabarty and Subhransu Maji.
\newblock The spectral bias of the deep image prior.
\newblock {\em arXiv preprint arXiv:1912.08905}, 2019.

\bibitem{chan2016plug}
Stanley~H Chan, Xiran Wang, and Omar~A Elgendy.
\newblock Plug-and-play admm for image restoration: Fixed-point convergence and
  applications.
\newblock {\em IEEE Transactions on Computational Imaging}, 3(1):84--98, 2016.

\bibitem{chandrasegaran2021closer}
Keshigeyan Chandrasegaran, Ngoc-Trung Tran, and Ngai-Man Cheung.
\newblock A closer look at fourier spectrum discrepancies for cnn-generated
  images detection.
\newblock In {\em Proceedings of the IEEE/CVF conference on computer vision and
  pattern recognition}, pages 7200--7209, 2021.

\bibitem{chen2020dip}
Yun-Chun Chen, Chen Gao, Esther Robb, and Jia-Bin Huang.
\newblock Nas-dip: Learning deep image prior with neural architecture search.
\newblock In {\em European Conference on Computer Vision}, pages 442--459.
  Springer, 2020.

\bibitem{cheng2019bayesian}
Zezhou Cheng, Matheus Gadelha, Subhransu Maji, and Daniel Sheldon.
\newblock A bayesian perspective on the deep image prior.
\newblock In {\em Proceedings of the IEEE/CVF Conference on Computer Vision and
  Pattern Recognition}, pages 5443--5451, 2019.

\bibitem{cortes1995support}
Corinna Cortes and Vladimir Vapnik.
\newblock Support-vector networks.
\newblock {\em Machine learning}, 20:273--297, 1995.

\bibitem{dabov2007video}
Kostadin Dabov, Alessandro Foi, and Karen Egiazarian.
\newblock Video denoising by sparse 3d transform-domain collaborative
  filtering.
\newblock In {\em 2007 15th European Signal Processing Conference}, pages
  145--149. IEEE, 2007.

\bibitem{darestani2021accelerated}
Mohammad~Zalbagi Darestani and Reinhard Heckel.
\newblock Accelerated mri with un-trained neural networks.
\newblock {\em IEEE Transactions on Computational Imaging}, 7:724--733, 2021.

\bibitem{durall2020watch}
Ricard Durall, Margret Keuper, and Janis Keuper.
\newblock Watch your up-convolution: Cnn based generative deep neural networks
  are failing to reproduce spectral distributions.
\newblock In {\em Proceedings of the IEEE/CVF conference on computer vision and
  pattern recognition}, pages 7890--7899, 2020.

\bibitem{frank2020leveraging}
Joel Frank, Thorsten Eisenhofer, Lea Sch{\"o}nherr, Asja Fischer, Dorothea
  Kolossa, and Thorsten Holz.
\newblock Leveraging frequency analysis for deep fake image recognition.
\newblock In {\em International conference on machine learning}, pages
  3247--3258. PMLR, 2020.

\bibitem{gandelsman2019double}
Yosef Gandelsman, Assaf Shocher, and Michal Irani.
\newblock " double-dip": unsupervised image decomposition via coupled
  deep-image-priors.
\newblock In {\em Proceedings of the IEEE/CVF Conference on Computer Vision and
  Pattern Recognition}, pages 11026--11035, 2019.

\bibitem{haralick1973textural}
Robert~M Haralick, Karthikeyan Shanmugam, and Its'~Hak Dinstein.
\newblock Textural features for image classification.
\newblock {\em IEEE Transactions on systems, man, and cybernetics},
  (6):610--621, 1973.

\bibitem{heckel2018deep}
Reinhard Heckel and Paul Hand.
\newblock Deep decoder: Concise image representations from untrained
  non-convolutional networks.
\newblock {\em arXiv preprint arXiv:1810.03982}, 2018.

\bibitem{heckel2019denoising}
Reinhard Heckel and Mahdi Soltanolkotabi.
\newblock Denoising and regularization via exploiting the structural bias of
  convolutional generators.
\newblock {\em arXiv preprint arXiv:1910.14634}, 2019.

\bibitem{ho2021neural}
Kary Ho, Andrew Gilbert, Hailin Jin, and John Collomosse.
\newblock Neural architecture search for deep image prior.
\newblock {\em Computers \& graphics}, 98:188--196, 2021.

\bibitem{jo2021rethinking}
Yeonsik Jo, Se~Young Chun, and Jonghyun Choi.
\newblock Rethinking deep image prior for denoising.
\newblock In {\em Proceedings of the IEEE/CVF International Conference on
  Computer Vision}, pages 5087--5096, 2021.

\bibitem{karras2021alias}
Tero Karras, Miika Aittala, Samuli Laine, Erik H{\"a}rk{\"o}nen, Janne
  Hellsten, Jaakko Lehtinen, and Timo Aila.
\newblock Alias-free generative adversarial networks.
\newblock {\em Advances in Neural Information Processing Systems}, 34:852--863,
  2021.

\bibitem{liu2019image}
Jiaming Liu, Yu Sun, Xiaojian Xu, and Ulugbek~S Kamilov.
\newblock Image restoration using total variation regularized deep image prior.
\newblock In {\em ICASSP 2019-2019 IEEE International Conference on Acoustics,
  Speech and Signal Processing (ICASSP)}, pages 7715--7719. Ieee, 2019.

\bibitem{liu2021swin}
Ze Liu, Yutong Lin, Yue Cao, Han Hu, Yixuan Wei, Zheng Zhang, Stephen Lin, and
  Baining Guo.
\newblock Swin transformer: Hierarchical vision transformer using shifted
  windows.
\newblock In {\em Proceedings of the IEEE/CVF international conference on
  computer vision}, pages 10012--10022, 2021.

\bibitem{loh2011classification}
Wei-Yin Loh.
\newblock Classification and regression trees.
\newblock {\em Wiley interdisciplinary reviews: data mining and knowledge
  discovery}, 1(1):14--23, 2011.

\bibitem{luisier2010image}
Florian Luisier, Thierry Blu, and Michael Unser.
\newblock Image denoising in mixed poisson--gaussian noise.
\newblock {\em IEEE Transactions on image processing}, 20(3):696--708, 2010.

\bibitem{mataev2019deepred}
Gary Mataev, Peyman Milanfar, and Michael Elad.
\newblock Deepred: Deep image prior powered by red.
\newblock In {\em Proceedings of the IEEE/CVF International Conference on
  Computer Vision Workshops}, pages 0--0, 2019.

\bibitem{odena2016deconvolution}
Augustus Odena, Vincent Dumoulin, and Chris Olah.
\newblock Deconvolution and checkerboard artifacts.
\newblock {\em Distill}, 1(10):e3, 2016.

\bibitem{rahaman2019spectral}
Nasim Rahaman, Aristide Baratin, Devansh Arpit, Felix Draxler, Min Lin, Fred
  Hamprecht, Yoshua Bengio, and Aaron Courville.
\newblock On the spectral bias of neural networks.
\newblock In {\em International Conference on Machine Learning}, pages
  5301--5310. PMLR, 2019.

\bibitem{roth2009fields}
Stefan Roth and Michael~J Black.
\newblock Fields of experts.
\newblock {\em International Journal of Computer Vision}, 82(2):205--229, 2009.

\bibitem{schwarz2021frequency}
Katja Schwarz, Yiyi Liao, and Andreas Geiger.
\newblock On the frequency bias of generative models.
\newblock {\em Advances in Neural Information Processing Systems},
  34:18126--18136, 2021.

\bibitem{shi2022measuring}
Zenglin Shi, Pascal Mettes, Subhransu Maji, and Cees~GM Snoek.
\newblock On measuring and controlling the spectral bias of the deep image
  prior.
\newblock {\em International Journal of Computer Vision}, 130(4):885--908,
  2022.

\bibitem{simoncelli2001natural}
Eero~P Simoncelli and Bruno~A Olshausen.
\newblock Natural image statistics and neural representation.
\newblock {\em Annual review of neuroscience}, 24(1):1193--1216, 2001.

\bibitem{soltanayev2018training}
Shakarim Soltanayev and Se~Young Chun.
\newblock Training deep learning based denoisers without ground truth data.
\newblock {\em Advances in neural information processing systems}, 31, 2018.

\bibitem{stein1981estimation}
Charles~M Stein.
\newblock Estimation of the mean of a multivariate normal distribution.
\newblock {\em The annals of Statistics}, pages 1135--1151, 1981.

\bibitem{ulyanov2018deep}
Dmitry Ulyanov, Andrea Vedaldi, and Victor Lempitsky.
\newblock Deep image prior.
\newblock In {\em Proceedings of the IEEE conference on computer vision and
  pattern recognition}, pages 9446--9454, 2018.

\bibitem{vaswani2017attention}
Ashish Vaswani, Noam Shazeer, Niki Parmar, Jakob Uszkoreit, Llion Jones,
  Aidan~N Gomez, {\L}ukasz Kaiser, and Illia Polosukhin.
\newblock Attention is all you need.
\newblock {\em Advances in neural information processing systems}, 30, 2017.

\bibitem{wang2021early}
Hengkang Wang, Taihui Li, Zhong Zhuang, Tiancong Chen, Hengyue Liang, and Ju
  Sun.
\newblock Early stopping for deep image prior.
\newblock {\em arXiv preprint arXiv:2112.06074}, 2021.

\bibitem{xu2018real}
Jun Xu, Hui Li, Zhetong Liang, David Zhang, and Lei Zhang.
\newblock Real-world noisy image denoising: A new benchmark.
\newblock {\em arXiv preprint arXiv:1804.02603}, 2018.

\bibitem{zhang2017beyond}
Kai Zhang, Wangmeng Zuo, Yunjin Chen, Deyu Meng, and Lei Zhang.
\newblock Beyond a gaussian denoiser: Residual learning of deep cnn for image
  denoising.
\newblock {\em IEEE transactions on image processing}, 26(7):3142--3155, 2017.

\bibitem{zhang2017learning}
Kai Zhang, Wangmeng Zuo, Shuhang Gu, and Lei Zhang.
\newblock Learning deep cnn denoiser prior for image restoration.
\newblock In {\em Proceedings of the IEEE conference on computer vision and
  pattern recognition}, pages 3929--3938, 2017.

\end{thebibliography}


Not Found
}

\end{document}